\documentclass[preprint,11pt]{aastex}

\begin{document}

\title{Multi-line Stokes inversion for prominence magnetic-field diagnostics}

\author{R. Casini,$^1$ A. L\'opez Ariste,$^2$ F. Paletou,$^3$ L.
L\'eger$^3$}

\affil{$^1$High Altitude Observatory,
National Center for Atmospheric Research,\altaffilmark{1}
P.O.~Box 3000, Boulder, CO 80307}

\altaffiltext{1}{The National Center for Atmospheric Research is
sponsored by the National Science Foundation.}

\affil{$^2$TH\'eMIS, CNRS UPS 853, C/\,V\'{\i}a L\'actea s/n, 
E-38200 La~Laguna, Tenerife, Spain}

\affil{$^3$Laboratoire d'Astrophysique de Toulouse-Tarbes, 
Universit\'e de Toulouse, CNRS, 14 av.~E.~Belin, 31400 Toulouse,
France}

\begin{abstract}
We present test results on the simultaneous inversion of the
Stokes profiles of the \ion{He}{1} lines at 587.6\,nm (D$_3$) and 
1083.0\,nm in prominences ($90^\circ$ scattering). We created 
datasets of synthetic Stokes profiles for the case of quiescent
prominences ($B<200$\,G), assuming a conservative value of $10^{-3}$ 
of the peak intensity for the polarimetric sensitivity of the simulated
observations. In this work, we focus on the error analysis for the 
inference of the magnetic field vector, under the usual assumption
that the prominence can be assimilated to a slab of finite optical
thickness with uniform magnetic and thermodynamic properties.
We find that the simultaneous inversion of the two lines significantly
reduces the errors on the inference of the magnetic field vector, with
respect to the case of single-line inversion. These results provide
a solid justification for current and future instrumental efforts 
with multi-line capabilities for the observations of solar prominences 
and filaments.
\end{abstract}

\maketitle

\section{Introduction}

The measurement of the vector magnetic field in solar prominences 
has become a prioritary goal to improve our understanding of the 
solar corona and its evolution. The long-term stability
of quiescent prominences (from several days to several weeks) suggests
that these structures of the solar atmosphere must be associated with a 
highly ordered topology of the magnetic field, right at the interface 
between the solar corona and the lower solar atmosphere. This
long-term stability is not easily disrupted, despite the fact that 
the visible structure of quiescent prominences appears to be continually 
affected by highly dynamical events, like rising large-scale voids
(\textit{bubbles}) and ascending and descending small-scale plumes
\citep{dT08,Be08}. This is rather strong evidence that the
stability of solar prominences---and its sudden disruption, when a
quiescent prominence eventually erupts, leading to a coronal mass
ejection (CME)---must somehow involve the magnetic topology of a 
much larger volume of the solar atmosphere than the actual visible 
structure, extending also to the prominence cavity and the corona above.
That is why a concerted effort for measuring the magnetic field 
vector in prominences and in the solar corona is fundamental
for the ultimate goal of understanding the manifestation of energetic 
events in the heliosphere, which are the main driver of space weather.

On the other hand, measurements of magnetic fields in prominences
and in the corona are difficult. They impose very strict requirements
on the instrumentation, and they require a deep understanding of the
mechanisms of formation of line scattering polarization in a
magnetized plasma. For this reason, only very recently the possibility
of performing routine measurements of magnetic fields in prominences
and in the corona has been given serious consideration, and new
anticipated large-scale instruments (ATST, EST, 
COSMO,\footnote{\texttt{http://www.cosmo.ucar.edu}} SOLAR-C) are being 
developed specifically with this goal in mind.

On the interpretational side, the number of lines that are
good diagnostics of prominence and coronal magnetic fields 
is quite restricted. For prominences, the primary choice has
converged, over the past two or three decades, to the two lines of 
\ion{He}{1} at 587.6\,nm (D$_3$) and 1083.0\,nm (10830). These two 
lines have been used at different times for studies of prominence 
magnetism \citep{Le77,Qu85,Pa01,TB02,Ca03,Me06,Ku09}. The long 
experience of the solar community in the magnetic inversion 
of photospheric lines has clearly demonstrated the advantage of 
multi-line polarimetric observations.
The MTR observing mode available at the TH\'eMIS telescope \citep{PM01}
is perfectly suited to accomplish such multi-line,
spectro-polarimetric observations.
Moreover, the implementation of optimized modulators, of new detectors
with better efficiency in the near-infrared, and the use of the grid
method of \cite{Se80}, allow since 2006 for the simultaneous and
co-spatial observations of the two multiplets of \ion{He}{1} in
prominences.
The HAO-NCAR Prominence Magnetometer \citep[ProMag;][]{El08}, which is
near completion, was specifically designed for this type of multi-line
diagnostics. 

It remains to prove that, for the specific case of the two \ion{He}{1} 
lines, such diverse-wavelength diagnostics of prominence magnetism
is indeed feasible, and that it would increase the reliability of the 
magnetic inversions. This is the motivation for
the study presented in this paper. The feasibility of such
multi-line approach to magnetic-field measurements in quiescent
prominences is demonstrated through its preliminary application to
simultaneous and co-spatial observations of these lines from TH\'eMIS.
Thereafter, in order to demonstrate the robustness of this diagnostics, 
we performed a statistical analysis of the inversion errors over 
several databases of simulated observations. This analysis clearly 
shows the significance of the improvement in magnetic-field inversion 
by using both \ion{He}{1} chromospheric lines.

\begin{figure}[t!]
\centering
\includegraphics[width=\hsize,clip]{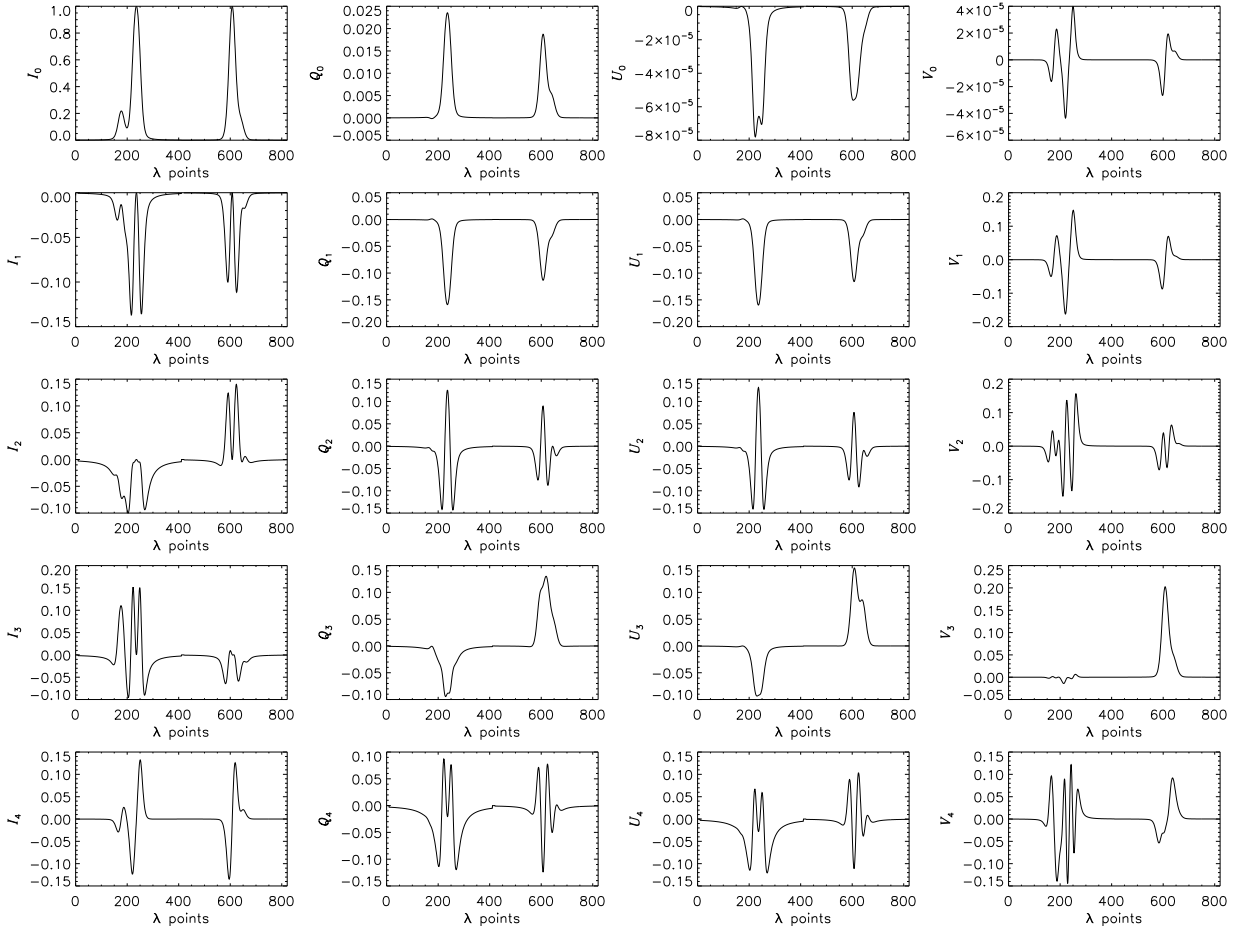}
\caption{\label{fig:eigenvectors}
First four PCA eigenprofiles (rows 2 to 5) for the four Stokes
parameters ($I$, $Q$, $U$, and $V$, from column 1 to 4, respectively) 
of the chromospheric lines \ion{He}{1} 10830 (left profiles) and D$_3$
(right profiles) formed in an environment typical of quiescent 
prominences. The first row shows
the mean Stokes profiles averaged over the entire parameter space 
spanned by our problem. The determination of the PCA eigenbasis from 
the merging of the spectral information of the two lines allows a 
clearer understanding of the physical correlations between the two 
lines with regards to the formation of their polarization signatures.}
\end{figure}

\section{PCA strategy for multi-line inversion}

The approach to magnetic inversion of line scattering polarization by
Principal Component Analysis (PCA) of the Stokes profiles has been 
described elsewhere \citep{LC02,Ca05}. The inversion is performed by
minimizing an appropriate Euclidean norm (\textit{PCA distance}),
which measures the deviation of the fitting profiles from the 
observations. This
minimization is performed in the ``dual space'' of the PCA coefficients,
which are the projections of the Stokes profiles onto a basis of 
\textit{eigenprofiles} previously determined for the problem at 
hand.\footnote{One of the determining characteristics of a basis 
of eigenprofiles for a given line formation problem is its 
completeness, that is, the 
property of representing the entire configuration space spanned by the
problem. In this sense, any PCA eigenbasis of Stokes profiles can be 
considered ``universal'' for the line formation problem at 
hand. The conditions for which one can trust the completeness 
of a PCA eigenbasis are the subject of ongoing research.} 
Since the PCA inversion is performed by searching for the best match 
of the observed profiles within a precomputed database of models, this 
method is very fast, and in addition is free from the risk of converging 
to local minima of the PCA distance.

Another advantage of the PCA approach to Stokes inversion is that 
all the essential spectro-polarimetric information about the line 
formation process, which is needed for a reliable inference of the 
magnetic and thermodynamic properties of the emitting plasma, is 
often encoded within a rather restricted set of eigenprofiles (typically 
between 3 and 6 components for each Stokes parameter, depending on the 
complexity of the line formation model). In fact,
the most significant (i.e., low-order) PCA eigenprofiles 
often present patterns that can be directly ascribed to specific 
physical mechanisms intervening in the line formation process
\citep{SL02}. Thus the inspection of the eigenprofiles can help 
identify which mechanisms are at work in a particular polarized line 
profile. Often this identification is obvious, but sometimes the 
eigenprofiles can reveal the presence of correlations between 
spectral signatures that may not be evident in the original line profiles. 

Figure \ref{fig:eigenvectors} shows the first few PCA eigenprofiles
for the two \ion{He}{1} chromospheric lines. These were computed after 
merging the spectral information from the two lines into a fictitious 
single spectrum. This allows us to reveal correlations of polarized 
spectral features between the two lines, which we can reasonably
expect must be present also in the line profiles of \ion{He}{1} 10830 and 
D$_3$ formed under realistic conditions, but which may not be as evident 
from a study of the separate sets of eigenprofiles for the two 
lines.

The four columns in Fig.~\ref{fig:eigenvectors} show the PCA components
of the Stokes parameters $I$, $Q$, $U$, and $V$, respectively. The
first row shows the mean profiles averaged over the entire parameter
space. As the magnetic field can attain all possible orientations, the
average $U$ and $V$ Stokes profiles tend to zero, as expected. In fact, 
the good approximation to zero of those signals, as illustrated by
Fig.~\ref{fig:eigenvectors}, is an indication of the good coverage of 
the parameter space provided by the original database of line profiles, 
which was used for the determination of the PCA eigenbasis. 
The successive rows give the first 4 orders of the PCA Stokes eigenvectors. 
Here we point out the most relevant features of these plots, which can 
give us a deeper insight in the formation mechanism of the two 
\ion{He}{1} chromospheric lines. The 1st-order, $Q$ and $U$ 
eigenvectors show that, to lowest order, the modifications of the
scattering linear polarization in the model (due to variations in the
radiation anisotropy, and to the magnetic Hanle effect) are positively
correlated between the two lines (i.e., both polarization increase or
decrease under the same physical mechanism). This fundamental trend is
only modified at higher order, as demonstrated by the 3rd-order
eigenvectors (4th row of plots) showing a negative correlation between 
the linear-polarization signatures of the two lines. 
At that same order, the Stokes $V$ eigenprofile shows the
unmistakable signature of atomic orientation in \ion{He}{1} D$_3$ (due
to the alignment-to-orientation mechanism; \citealt{La82,Ke84}). It
is interesting to note that almost no signal is present for
\ion{He}{1} 10830 in such case. In other words, if a physical condition
existed in a quiescent prominence, such that the resulting Stokes $V$
of \ion{He}{1} D$_3$ were completely due to atomic orientation, then
one should expect to see a negligibly small Stokes $V$ signal for the
\ion{He}{1} 10830 line formed in the same region of the prominence.
Therefore, the simultaneous and co-spatial observation of such a
peculiar Stokes $V$ profile in \ion{He}{1} D$_3$ {\em and} of a sizable 
(Zeeman) Stokes $V$ profile in \ion{He}{1} 10830 would imply that the
formation regions for the two lines cannot be the same, a possibility 
that currently is not contemplated by our inversion model. So the
3rd-order, Stokes-$V$ eigenprofile shown in
Fig.~\ref{fig:eigenvectors} provides an interesting proxy of
non-standard line formation scenarios for the chromospheric lines of
\ion{He}{1}. On the other hand, the 1st-order, Stokes-$V$ eigenprofile 
shows that the dominant contribution to Stokes $V$ for both lines, in
our model, is in fact due to the Zeeman effect. This is easily explained 
by the fact that, in our parameter space, we considered field strengths 
up to 200\,G, which are well beyond the level-crossing regime for 
\ion{He}{1} D$_3$ (30--40\,G), at which the alignment-to-orientation 
mechanism is most effective in the $3{}^3$D term of \ion{He}{1}.

In extending our PCA code \citep{LC02} to multi-line inversion, 
the database creation was modified so that each line's database is 
calculated for 
exactly the same set of magnetic and plasma models. This decision was 
made in order to minimize the effects of the discrete nature of the PCA
database,\footnote{In principle, in the case of infinitely dense 
databases, it would not matter if the databases for two different lines 
had been calculated on two completely independent set of models.} and 
also to eliminate inversion artifacts due to the presence of
ambiguous magnetic configurations \citep[see, e.g.,][]{La82}.
We then introduced a generalized PCA distance
\begin{equation}	\label{eq:distance}
\bar{d^2}=\sum_{i=1}^N w_i d_i^2\;,
\end{equation}
where $N$ is the total number of spectral lines in the model 
\citep[for the currently adopted model of \ion{He}{1}, $N=6$; see][]{LC02}, 
$d_i$ is the PCA distance 
for the $i$-th line, and $w_i$ is a factor ranging between 0 and 1. This 
last quantity (which at this time is being fixed before the inversion) 
allows to switch on and off the inversion of a given line, or to give
different weights to the lines that are being inverted. The
simultanoues PCA inversion of multiple spectral lines thus requires the
minimization of the generalized PCA distance defined by
eq.~(\ref{eq:distance}).

One could alternatively use the eigenprofiles of 
Fig.~\ref{fig:eigenvectors} for the inversion of simultaneous 
spectro-polarimetric data in the two chromospheric lines of \ion{He}{1}, 
with the advantage of dealing with only one database of PCA coefficients 
and one PCA distance for both lines. Since the quality of the inversion 
is independent of the choice of the eigenprofile basis, this alternate 
approach is totally legitimate. On the other hand, this would require 
the preliminary treatment of all spectro-polarimetric data to merge the 
information from the two lines into fictitious Stokes profiles analogous 
to those shown in Fig.~\ref{fig:eigenvectors}. An important advantage in 
keeping separate databases for each line is that it easily allows for the 
testing of different combinations of lines for magnetic inversion. This 
is the main argument in favor of our approach to multi-line PCA inversion 
based on the minimization of a multi-line PCA distance as given by 
eq.~(\ref{eq:distance}).

\begin{figure}[t!]
\centering
\includegraphics[height=.88\vsize,clip]{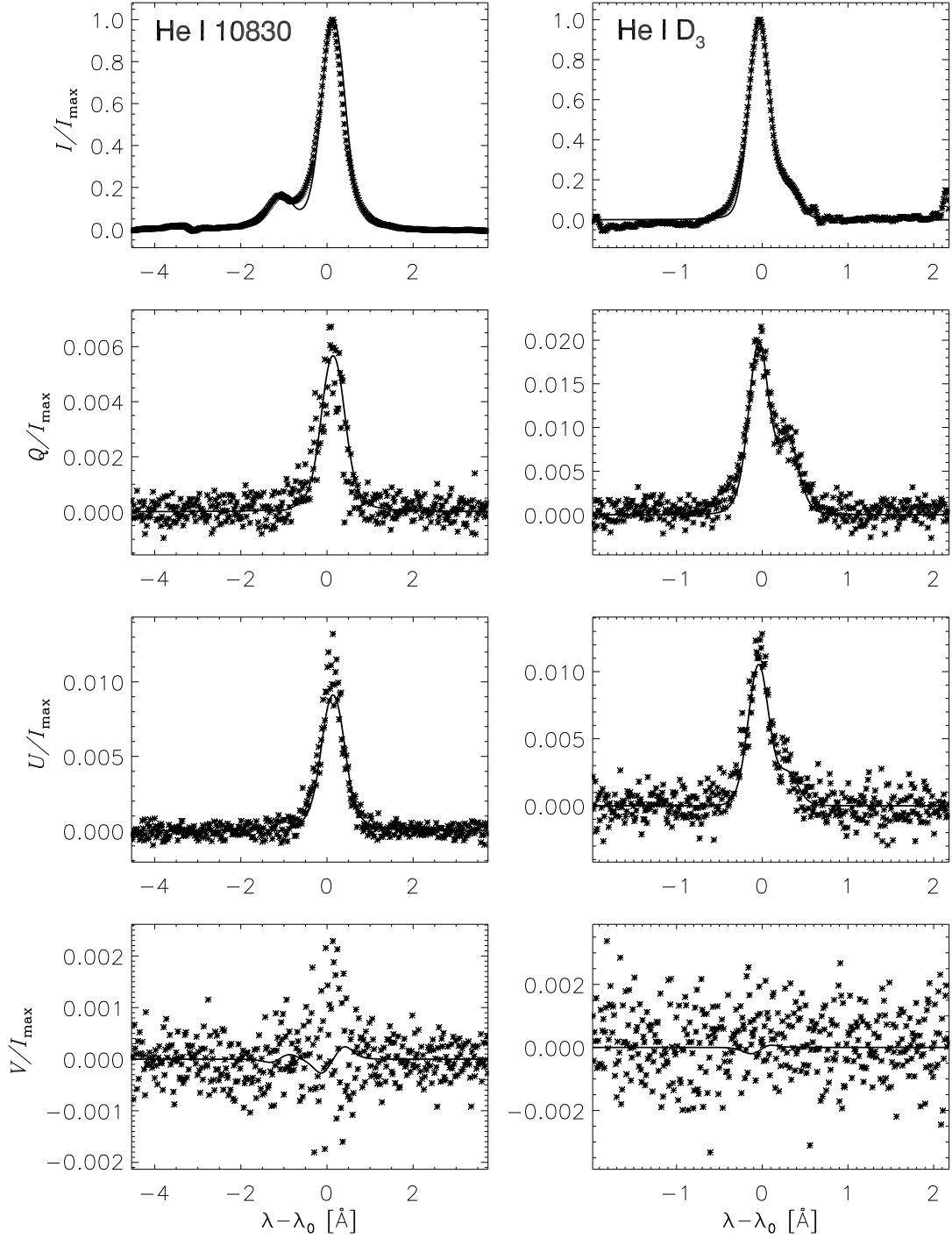}\vskip 15pt
\caption{\label{fig:THeMIS}
Multi-line inversion of simultaneous and co-spatial spectro-polarimetric 
observations of \ion{He}{1} 10830 (left) and D$_3$ (right) in a
quiescent prominence, taken with TH\'eMIS on June 29, 2007. 
The inverted vector magnetic field for this example is $B=3.0\,\rm G$,
$\vartheta_B=57.8^\circ$, $\varphi_B=42.7^\circ$.}
\end{figure}

We preliminarly applied our multi-line PCA inversion code to 
simultaneous and co-spatial observations of the two chromospheric 
\ion{He}{1} lines in a quiescent prominence, which were acquired in 
June 2007 at TH\'eMIS (see Fig.~\ref{fig:THeMIS}). This application 
to only a few spatial points of the prominence has obviously no 
relevance for an improved understanding of the magnetic topology of 
these solar structures. However, the successful fit of these
simultaneous, multi-line observations marks an important advance in the
spectro-polarimetric diagnostics of scattering polarization, 
confirming the feasibility of multi-line inversion for magnetic 
studies of solar prominences. This result strongly advocates for 
a consistent design of future solar instrumentation that allow
multi-line observations of chromospheric lines at the solar limb.

\begin{table}
\centering
\begin{tabular}{rrrccccccc}
\hline\hline
\noalign{\vskip 5pt}
\multispan{3}
	\hfill inverted line(s):
	& \multispan{2}\hfill 10830 \hfill
	& \multispan{2}\hfill D$_3$ \hfill
	& \multispan{2}\hfill 10830 + D$_3$ \hfill\\
\noalign{\vskip 5pt}
\multispan{3}
	& \multispan{2}\hfill $\overbrace{\hbox to 60pt{}}$ \hfill
	& \multispan{2}\hfill $\overbrace{\hbox to 60pt{}}$ \hfill
	& \multispan{2}\hfill $\overbrace{\hbox to 60pt{}}$ \hfill\\
$B_{\rm min}$ & $B_{\rm max}$ 
	& $\Delta X$
	& 50\% & 90\% & 50\% & 90\% & 50\% & 90\% \\
\noalign{\vskip 3pt}
\hline\hline
\noalign{\vskip 3pt}
\multispan{2} & $B\,\rm(G)$ 
	& 22.5 & 97.3 & 12.3 & 71.1 & 10.3 & 57.0 \\
\noalign{\vskip 3pt}
0\,G&200\,G   & $\Theta_B\,(^\circ)$
	& 6.1 & 25.2 & 4.4 & 18.1 & 3.2 & 12.7 \\
\noalign{\vskip 3pt}
\multispan{2} & $\Phi_B\,(^\circ)$
	& 5.6 & 179.0 & 6.0 & 178.7 & 3.3 & 179.1 \\
\noalign{\vskip 3pt}
\hline
\noalign{\vskip 3pt}
\multispan{2} & $B\,\rm(G)$
	& 4.0 & 99.8 & 5.6 & 38.2 & 2.3 & 32.5 \\
\noalign{\vskip 3pt}
0\,G&10\,G & $\Theta_B\,(^\circ)$ 
	& 9.1 & 32.4 & 10.0 & 33.9 & 6.1 & 23.4 \\
\noalign{\vskip 3pt}
\multispan{2} & $\Phi_B\,(^\circ)$
	& 15.7 & 177.8 & 32.4 & 174.6 & 8.6 & 177.8 \\
\noalign{\vskip 3pt}
\hline
\noalign{\vskip 3pt}
\multispan{2} & $B\,\rm(G)$
	& 32.1 & 83.3 & 23.9 & 70.9 & 18.8 & 68.2 \\
\noalign{\vskip 3pt}
100\,G&110\,G & $\Theta_B\,(^\circ)$ 
	& 5.7 & 25.2 & 4.1 & 16.9 & 3.2 & 11.2 \\
\noalign{\vskip 3pt}
\multispan{2} & $\Phi_B\,(^\circ)$
	& 5.4 & 179.0 & 3.7 & 178.9 & 2.6 & 179.0 \\
\noalign{\vskip 3pt}
\hline\hline
\end{tabular} 
\caption{\label{tab:errors}
Comparison of the errors at the 50\% and 90\% confidence level 
on the inferred values of $B$, $\Theta_B$, and $\Phi_B$,  for single-line
and multi-line inversions.}
\end{table}

\begin{figure}[t!]
\includegraphics[height=.27\vsize]{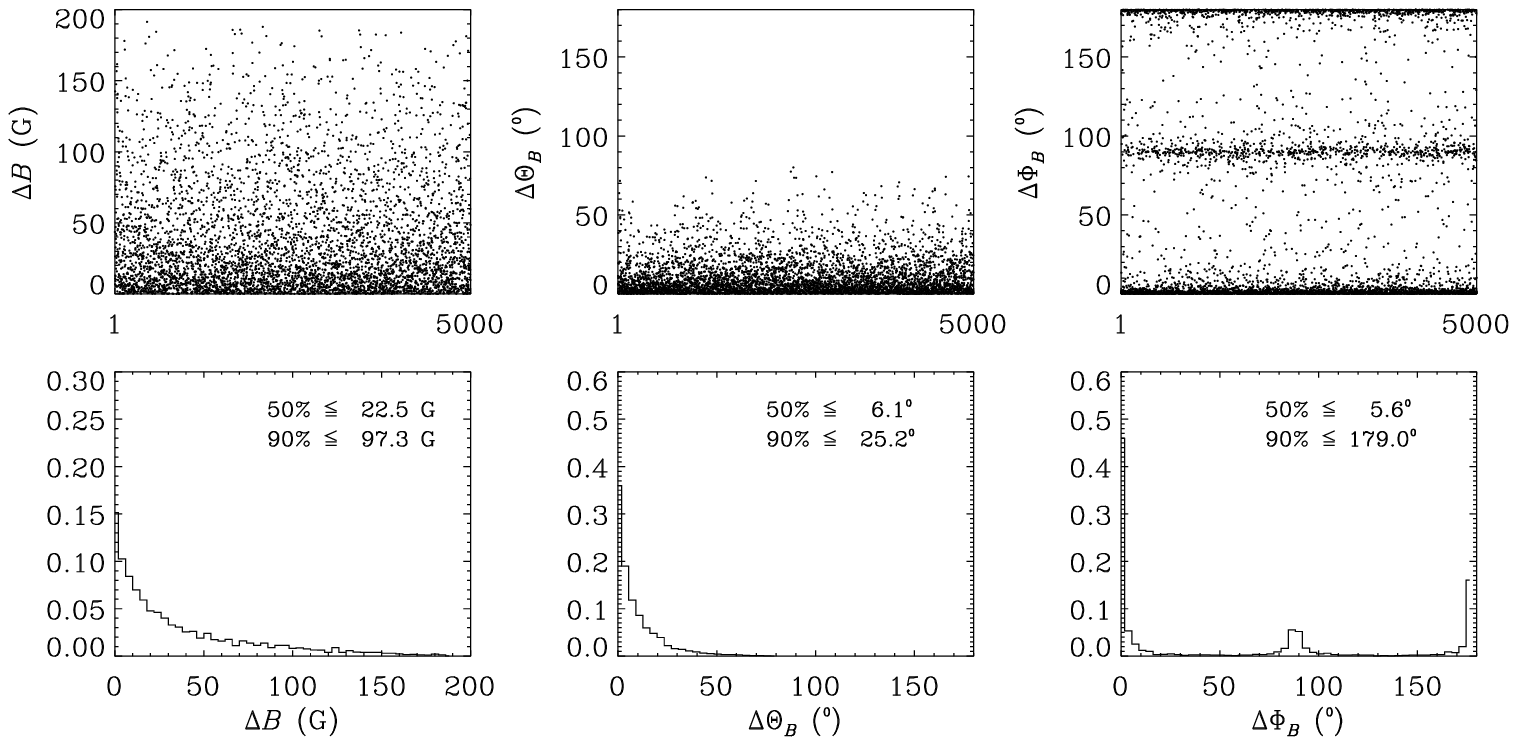}
	\raise .25\vsize\hbox{\ion{He}{1} 10830}\hfill\vspace{5pt}
\includegraphics[height=.27\vsize]{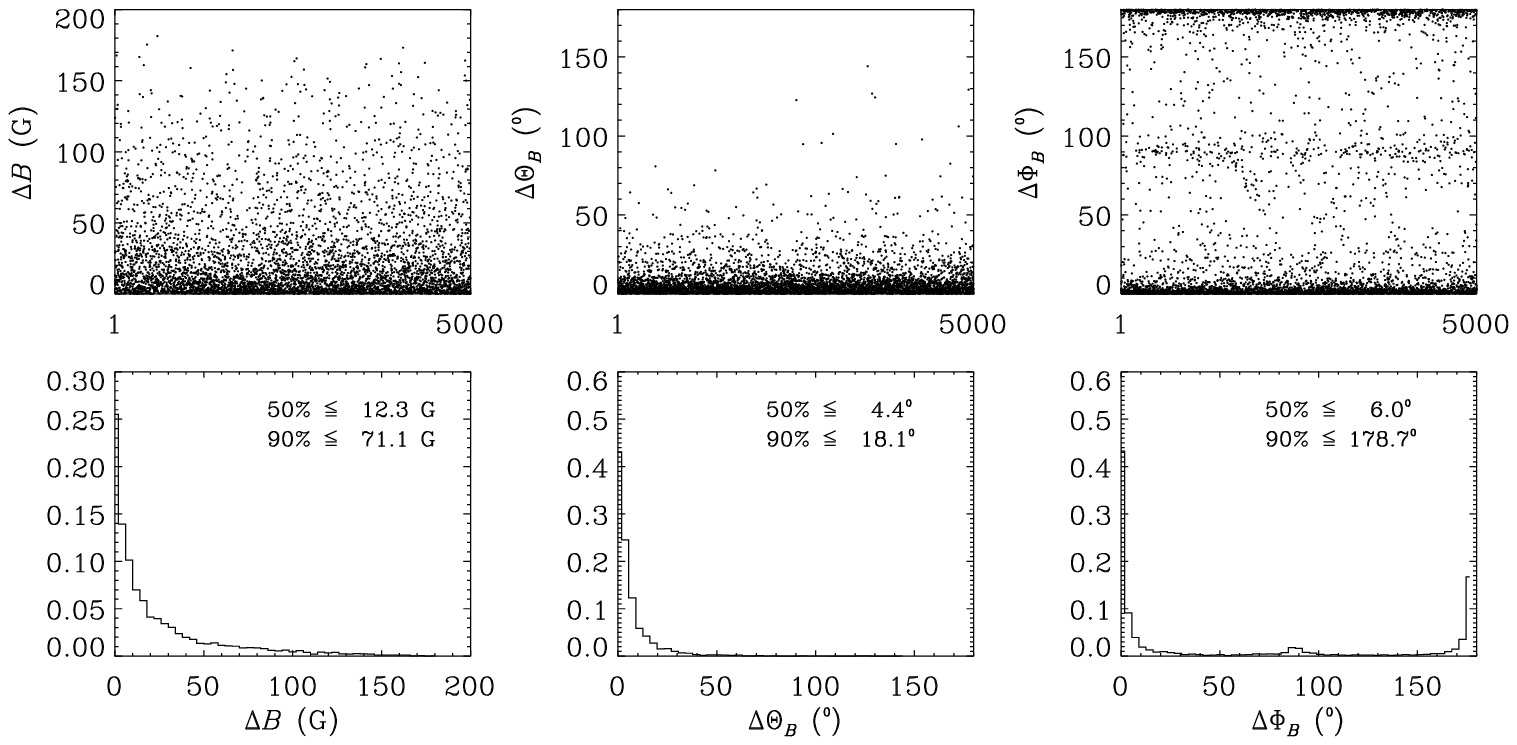}
	\raise .25\vsize\hbox{\ion{He}{1} D$_3$}\hfill\vspace{5pt}
\includegraphics[height=.27\vsize]{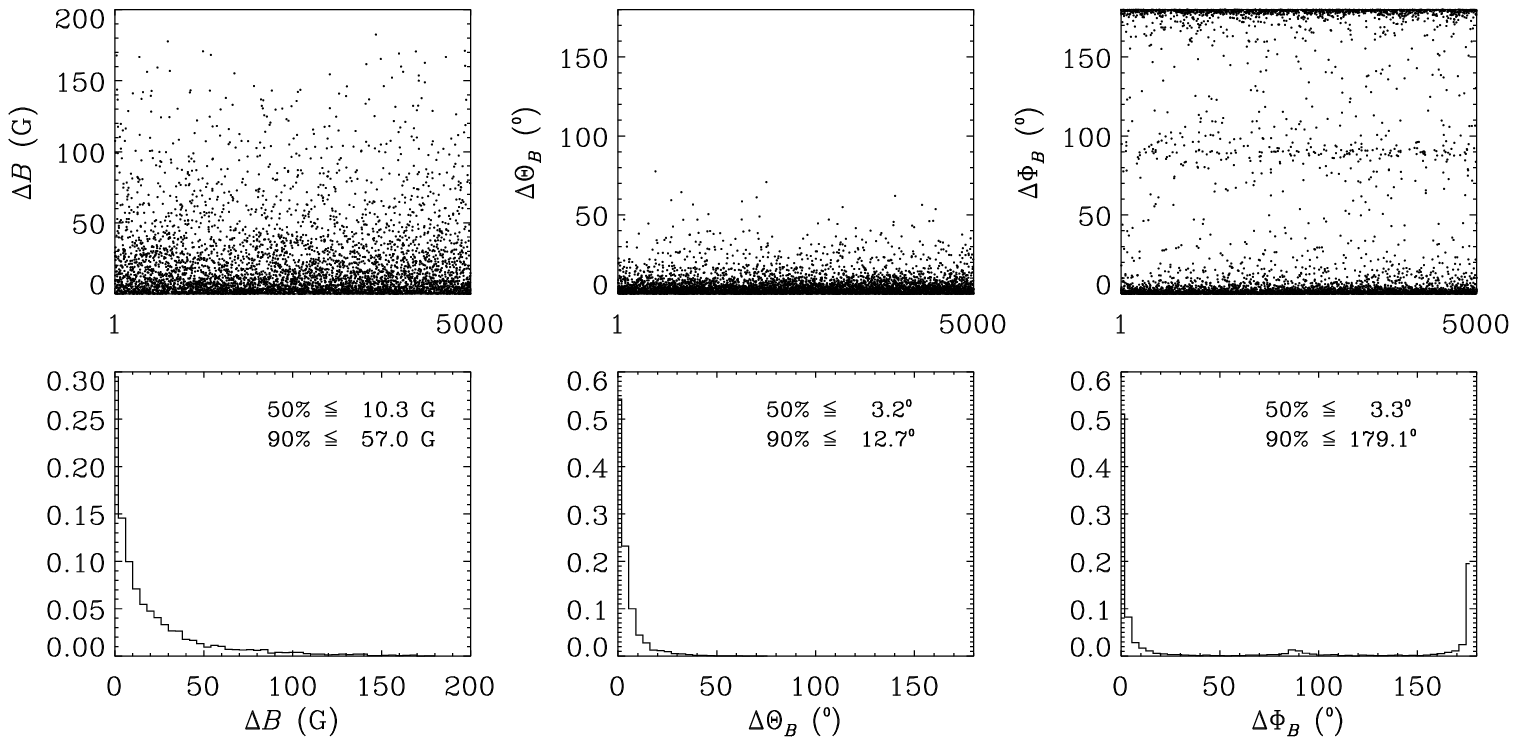}
	\raise .25\vsize\hbox{\ion{He}{1} 10830 + D$_3$}\hfill
\caption{\label{fig:full_range}
Scatter-plot and histogram distributions of the inversion errors on
magnetic field strength, $B$, inclination of the magnetic field 
along the LOS, $\Theta_B$, and position angle of the magnetic field
projection on the POS, $\Phi_B$. The database of 5000 synthetic
observations spans the entire range of magnetic strength used in the
construction of the inversion database with 150000 models. \textit{Top two
rows:} inversion of \ion{He}{1} 10830; \textit{middle
two rows:} inversion of \ion{He}{1} D$_3$; \textit{bottom
two rows:} simultaneous inversion of both lines.}
\end{figure}

\begin{figure}[t!]
\includegraphics[height=.27\vsize]{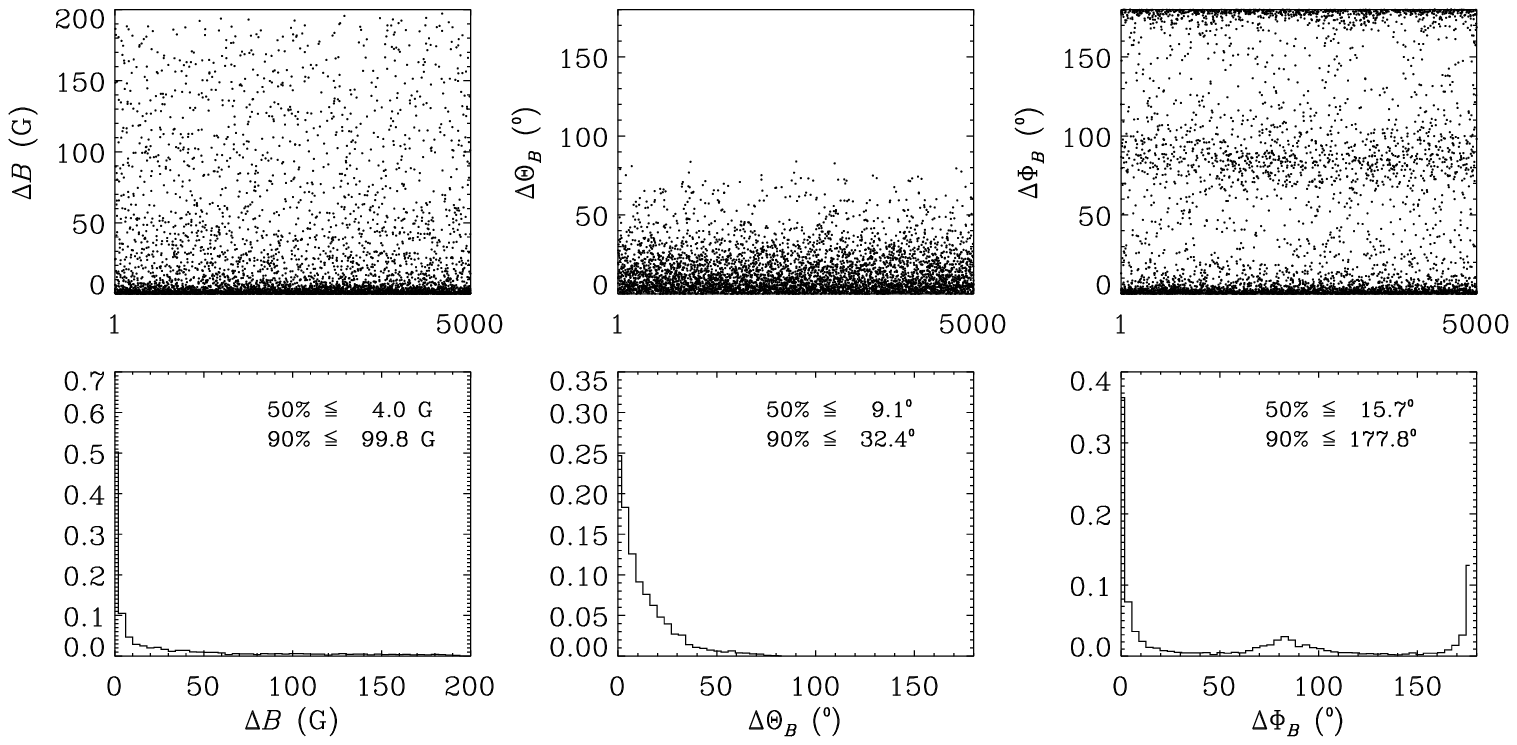}
	\raise .25\vsize\hbox{\ion{He}{1} 10830}\hfill\vspace{5pt}
\includegraphics[height=.27\vsize]{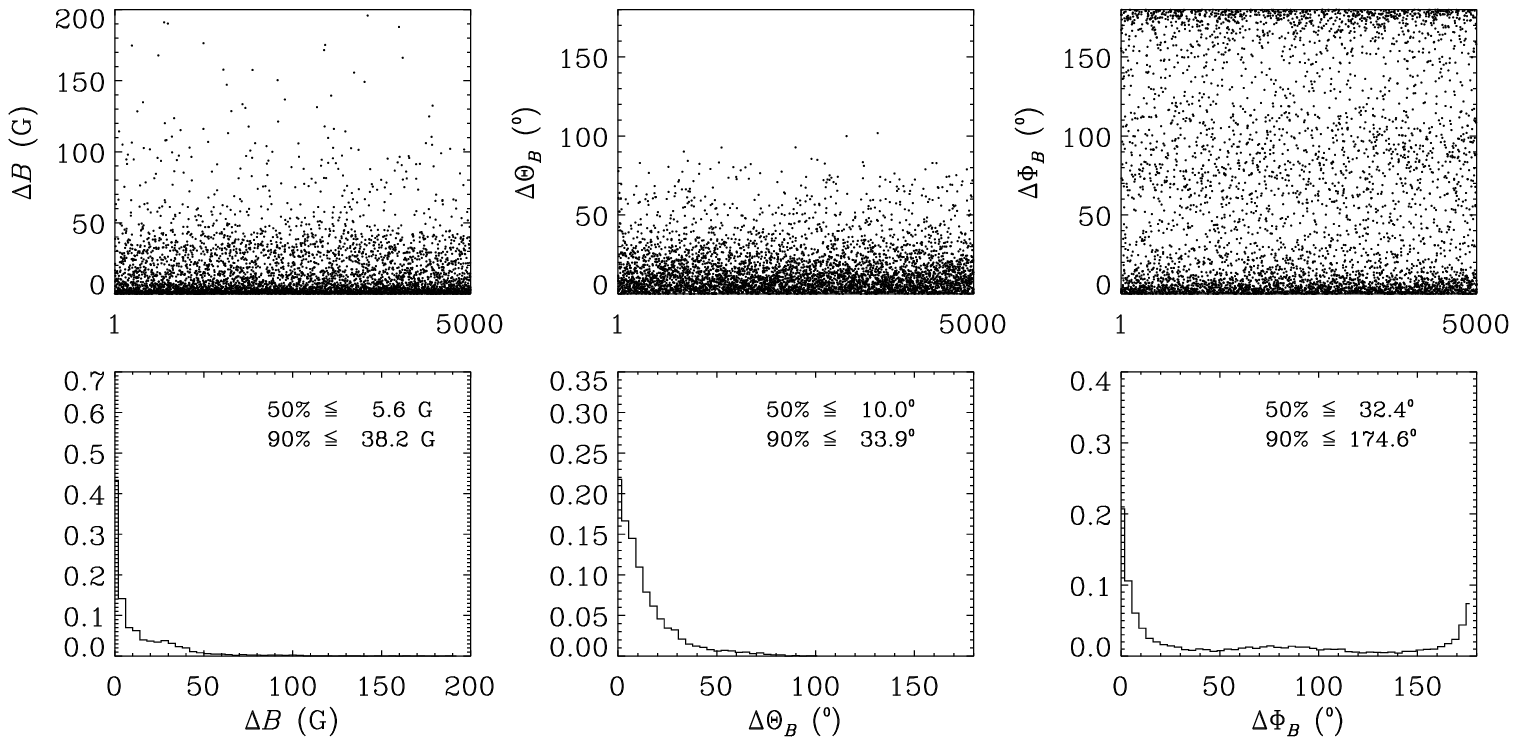}
	\raise .25\vsize\hbox{\ion{He}{1} D$_3$}\hfill\vspace{5pt}
\includegraphics[height=.27\vsize]{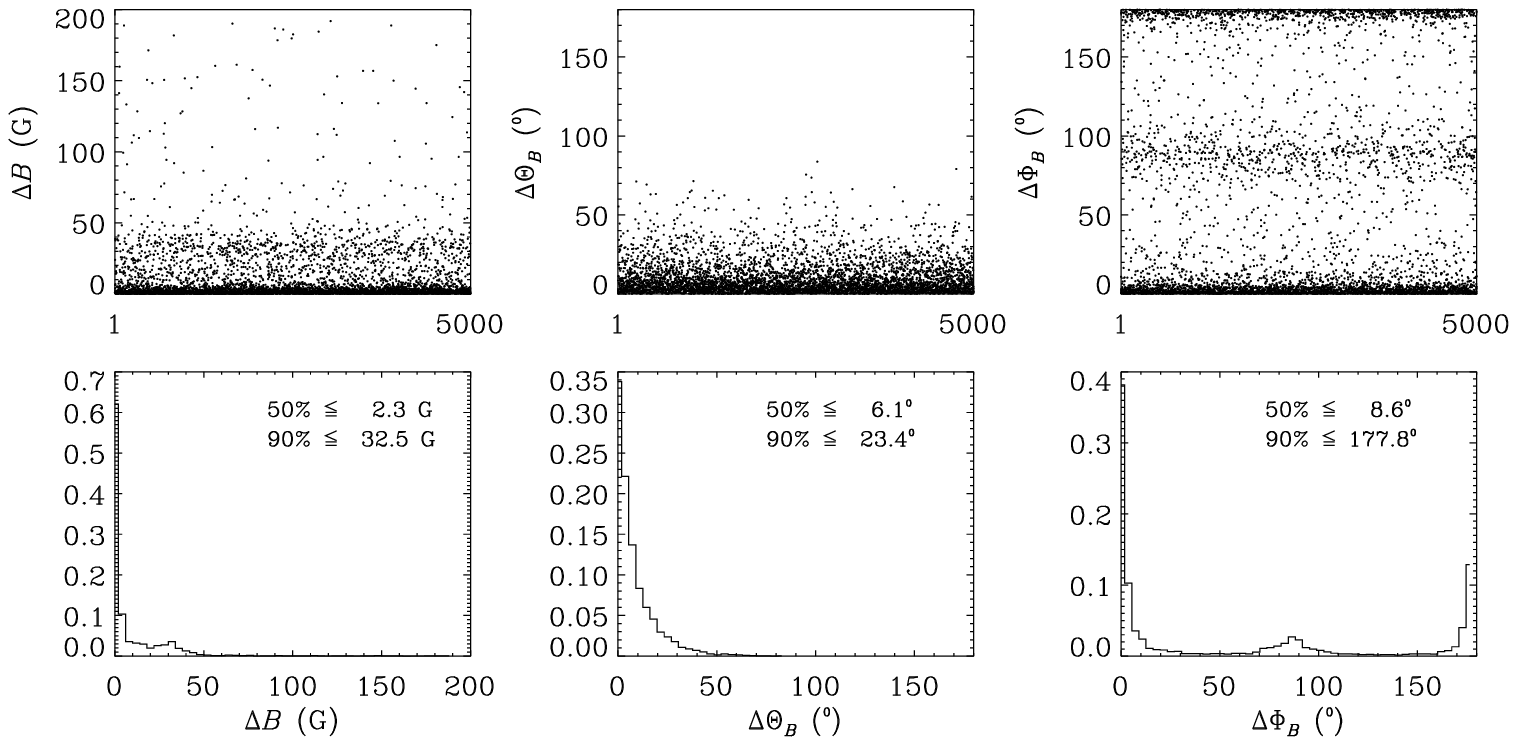}
	\raise .25\vsize\hbox{\ion{He}{1} 10830 + D$_3$}\hfill
\caption{\label{fig:lowB}
Same as Fig.~\ref{fig:full_range}, but for a database of simulated 
observations spanning weak field strengths between 0 and 10\,G.}
\end{figure}

\begin{figure}[t!]
\includegraphics[height=.27\vsize]{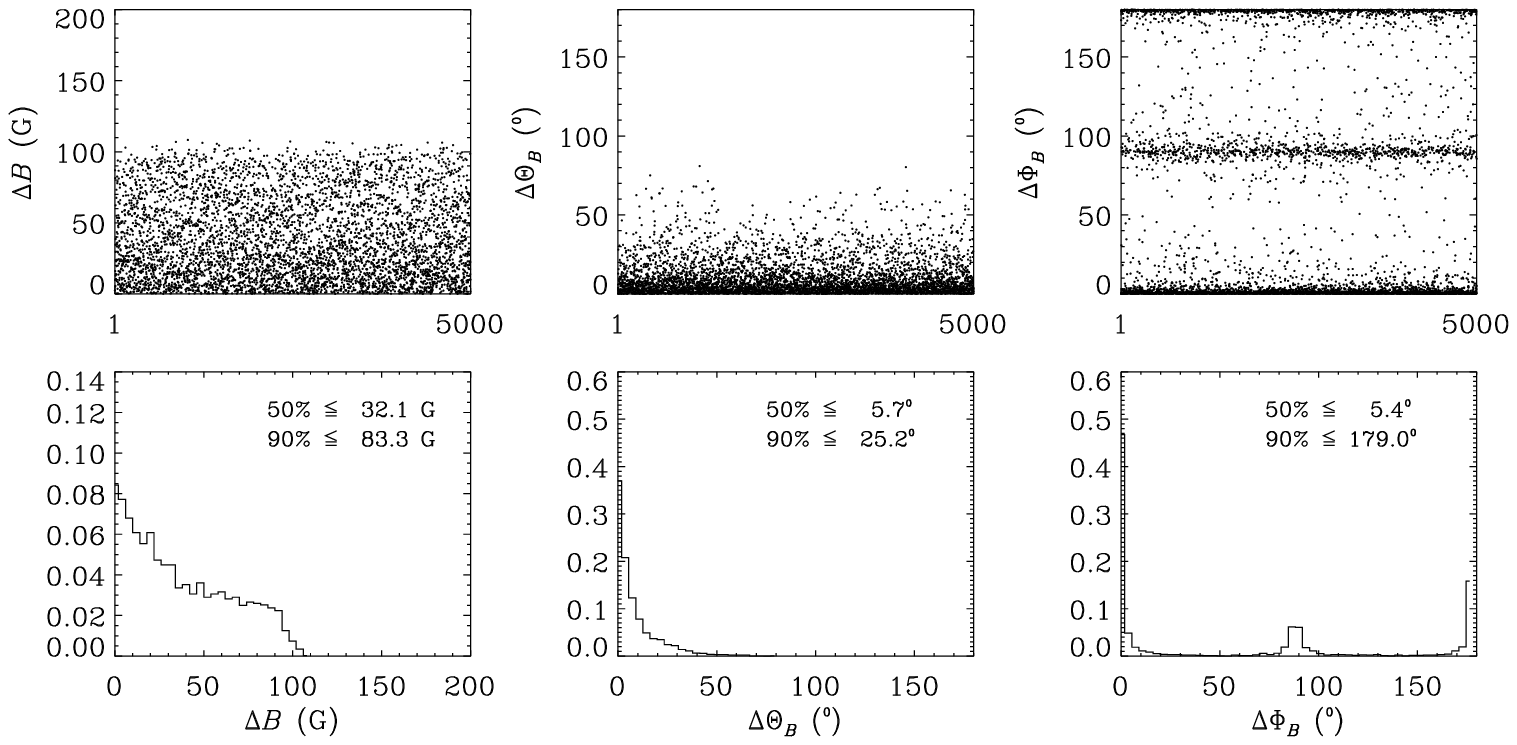}
	\raise .25\vsize\hbox{\ion{He}{1} 10830}\hfill\vspace{5pt}
\includegraphics[height=.27\vsize]{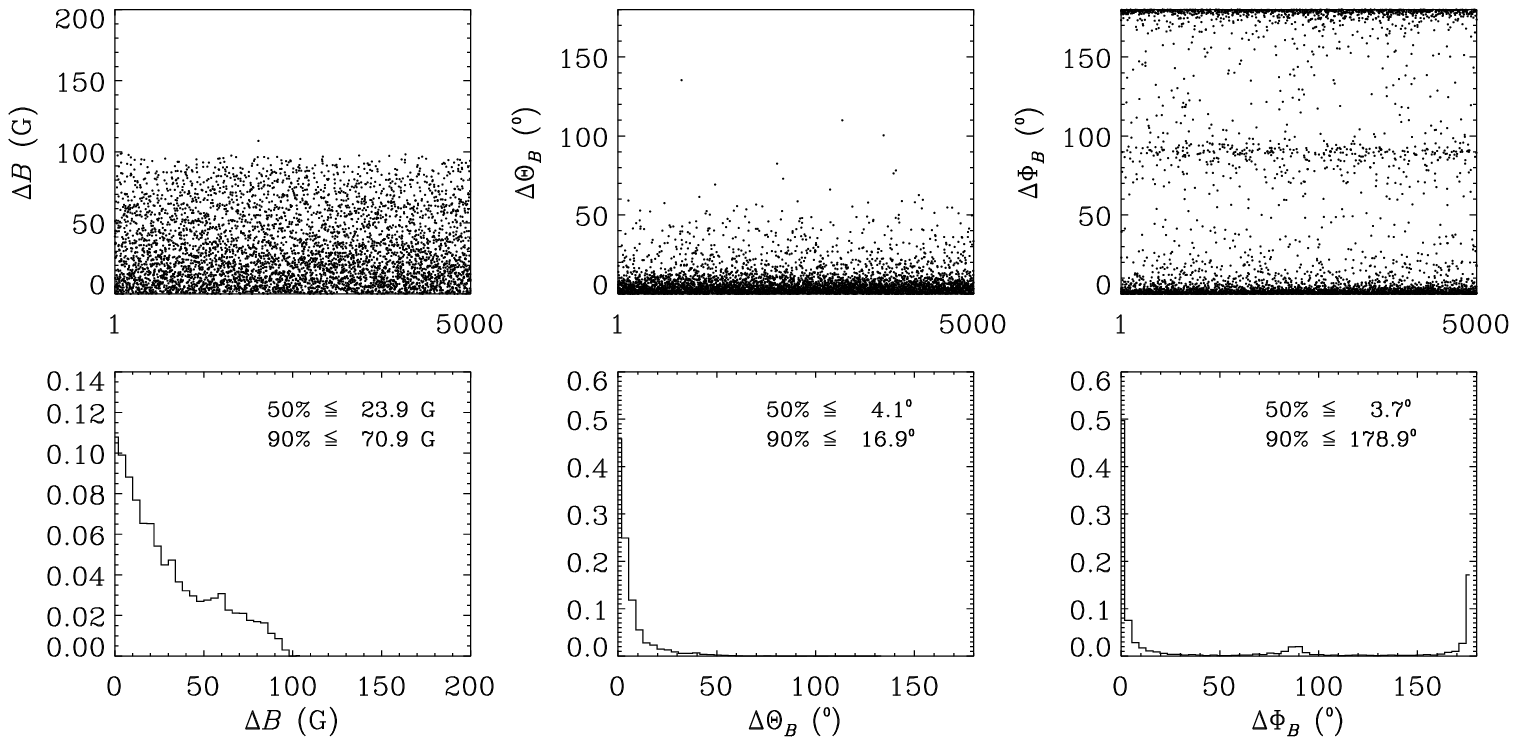}
	\raise .25\vsize\hbox{\ion{He}{1} D$_3$}\hfill\vspace{5pt}
\includegraphics[height=.27\vsize]{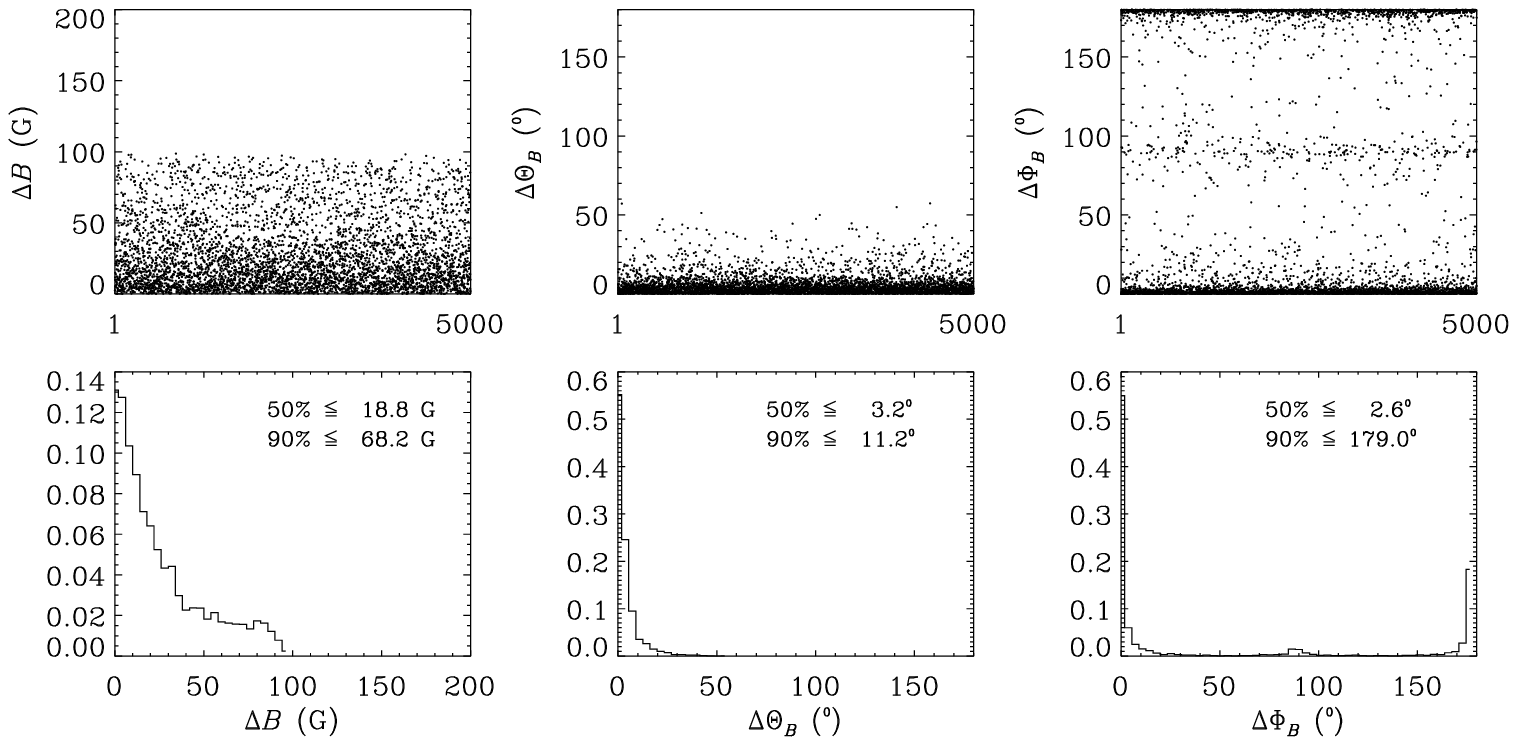}
	\raise .25\vsize\hbox{\ion{He}{1} 10830 + D$_3$}\hfill
\caption{\label{fig:highB}
Same as Fig.~\ref{fig:full_range}, but for a database of simulated
observations spanning medium field strengths between 100 and 110\,G.}
\end{figure}

\section{Test results}

We considered the case of quiescent prominence observations occurring
between heights of 0.01 and 0.06\,$R_\odot$ above the solar limb. The
inclination of the line of sight with respect to the radial direction
through the prominence was restricted between $85^\circ$ and 
$95^\circ$ in order to avoid the appearence of mixed prominence/filament 
cases in the synthetic database. The magnetic field strength was limited to
$B<200$\,G, without any restriction on the field direction. Finally we
assumed a plasma temperature (including micro-turbulence) ranging
between 10000 and 15000\,K, and an optical depth at line center
for the prominence slab ranging between 0.5 and 1.5 for \ion{He}{1}
10830. With these
parameter intervals, we computed a PCA database of 250\,000 models. 
For the parameter space described above, such a large size of the 
database is needed in order to mitigate the contribution to the 
inversion errors coming from the discrete nature of the database. 
So we can anticipate that the inversion errors in our tests are 
dominated by the intrinsic uncertainties (ambiguities) in the 
polarized line formation of the \ion{He}{1} lines in prominences.

Next we created three different datasets of synthetic observations, 
each with 5000 models, to be inverted against the PCA database. One 
of the datasets spans the same parameter ranges given above. The two 
other datasets cover a subset of magnetic field strengths, sampling 
respectively weak fields ($B<10$\,G) and medium fields 
($100\,{\rm G}<B<110\,\rm G$). 

Figure~\ref{fig:full_range} shows the scatter plots and
histograms of the errors on the inferred magnetic field vector, 
expressed in the reference system of the observer, for the
first dataset. The top set of graphs shows the inversion results 
using only \ion{He}{1} 10830, the second set using only
\ion{He}{1} D$_3$, and the bottom set using both lines equally
weighted ($w_i=1$). It is evident that, on average, 
\ion{He}{1} D$_3$ performs significantly better than \ion{He}{1} 
10830 in inferring the vector magnetic field information over 
the entire parameter space. However, \ion{He}{1} 10830 helps 
in reducing the inversion errors, particularly on the field 
direction. For the histogram, we report the widths of the error 
distribution containing 50\% and 90\% of the inverted models, 
respectively.

Figures~\ref{fig:lowB} and \ref{fig:highB} show similar results for
the inversion of the two observation databases created on a 
restricted range (10\,G wide) of magnetic field strengths. We see that 
\ion{He}{1} 10830 provides most of the information that is 
needed for the inference of weak fields (Fig.~\ref{fig:lowB}). 
Instead, \ion{He}{1} D$_3$ carries more information for larger field
strength, as indicated by the results of Fig.~\ref{fig:highB}. This is
in general agreement with the fact that the critical field strength
for the Hanle effect increases by nearly an order of magnitude going
from \ion{He}{1} 10830 to \ion{He}{1} D$_3$, where \ion{He}{1} 
10830 is sensitive to fields between approximately 0.1 and 
10\,G. It is important to notice that, for all three cases depicted 
in Figures~\ref{fig:full_range}--\ref{fig:highB}, the simultaneous
inversions of the two lines give consistently the smallest errors 
at both the 50\% and 90\% confidence levels. All the errors shown in
the histograms of Figures~\ref{fig:full_range}--\ref{fig:highB} are 
summarized in Table~\ref{tab:errors}. 

In all the inversions, the 90\% error on the position angle of the 
magnetic field projection on the plane of the sky, $\Delta\Phi_B$, 
is always dominated by errors close to $180^\circ$, because of the 
well-known $180^\circ$ ambiguity of polarization measurements. We 
can very clearly distinguish the peaks at $90^\circ$ in the 
top set of panels of Figs.~\ref{fig:full_range} and \ref{fig:highB}. 
These are determined by the behavior of resonance scattering 
polarization in the asymptotic regime of the Hanle effect for 
\ion{He}{1} 10830 \citep{Ca05}. \ion{He}{1} D$_3$ is much less 
affected by such $90^\circ$ ambiguity, for this specific range of 
field strengths. However, it is interesting to note how the 
simultaneous inversion of \ion{He}{1} 10830 and D$_3$ further 
reduces the probability of inversion errors associated with the 
$90^\circ$ ambiguity.

 \section{Conclusions}

The complexity of magnetic diagnostics in solar prominences,
where the observed polarization is dominated by scattering processes, 
poses particularly strong demands on the robustness of the inversion 
strategy. One would greatly profit from using multi-line diagnostics,
but the feasibility and reliability of such an approach in the case of
scattering-dominated, non-LTE, radiative transfer had yet to be
demonstrated. In this paper, we took on this task, and considered in
particular the simultaneous inversion of the chromospheric \ion{He}{1}
lines at 587.6\,nm and 1083.0\,nm.

We performed a statistical study of magnetic-field inference in solar 
prominences through PCA-based, multi-line inversion, as applied to 
simultaneous observations of the \ion{He}{1} D$_3$ and 10830. The 
statistical analysis of the
inversion errors on the inferred vector magnetic field was conducted
on a database of simulated observations of the two lines. Our analysis
confirmed that \ion{He}{1} D$_3$ carries the greatest diagnostic
content for typical average fields of quiescent prominences 
($B\sim 10$\,G and higher), but also demonstrated that the added
information carried by the polarization signatures of \ion{He}{1}
10830 significantly improves the determination of the magnetic field
geometry. On the other hand, \ion{He}{1} 10830 is fundamental for the
vector measurement of weak magnetic field ($B<10$\,G), although the 
use of both lines is even more important for the overall reduction of 
the inversion errors on both field strength and geometry in this case.

We also demonstrated the applicability of such diagnostics to real
data, by inverting some recent simultaneous and co-spatial observations 
of a quiescent prominence, which were taken with TH\'eMIS in the two 
\ion{He}{1} chromospheric lines and in H$\alpha$. Although we did not 
put any emphasis on the interpretation of those observations, the fact 
that real data are indeed amenable to such multi-line diagnostics allows 
us to extends the results of our error analysis of multi-line PCA 
inversion to future observations of quiescent prominences.

\acknowledgments We wish to thank A.~Asensio Ramos (IAC) for insightful
discussions on various statistical aspects of line inversion. B.~ Lites
(HAO) is acknowledged for carefully reading the manuscript and for
helpful comments that have improved the presentation of this work.
TH\'eMIS is operated by CNRS-CNR at the Observatorio del  Teide of 
the Instituto de Astrof\'{\i}sica de Canarias (Tenerife, Spain).

\end{document}